\documentclass[journal,onecolumn]{IEEEtran}
\usepackage{authblk}
\usepackage{graphicx}
\usepackage{amsfonts}
\usepackage{subcaption}
\usepackage{amssymb}
\usepackage{setspace}
\usepackage{cite}
\usepackage{amsmath}

\ifCLASSINFOpdf

\else

\fi

\begin{document}

\title{Coded Federated Computing in Wireless Networks with Straggling Devices and Imperfect CSI}

\author{Sukjong~Ha\IEEEauthorrefmark{1}, Jingjing~Zhang\IEEEauthorrefmark{2}, Osvaldo~Simeone\IEEEauthorrefmark{2},      and~Joonhyuk~Kang\IEEEauthorrefmark{1}
}
\affil{\IEEEauthorrefmark{1} Korea Advanced Institute of Science and Technology, School of Electrical Engineering, South Korea\\
\IEEEauthorrefmark{2} King's College London, Centre for Telecommunications Research, London, United Kingdom\\
\IEEEauthorrefmark{1} sj.ha@kaist.ac.kr, jhkang@ee.kaist.ac.kr, \IEEEauthorrefmark{2} \{jingjing.1.zhang, osvaldo.simeone\}@kcl.ac.uk}


\maketitle
\begin{abstract}
Distributed computing platforms typically assume the availability of reliable and dedicated connections among the processors. 
This work considers an alternative scenario, relevant for wireless data centers and federated learning, in which the distributed processors, operating on generally distinct coded data, are connected via shared wireless channels accessed via full-duplex transmission. 
The study accounts for both wireless and computing impairments, including interference, imperfect Channel State Information, and straggling processors, and it assumes a Map-Shuffle-Reduce coded computing paradigm. 
The total latency of the system, obtained as the sum of computing and communication delays, is studied for different shuffling strategies revealing the interplay between distributed computing, coding, and cooperative or coordinated transmission.
\end{abstract}

\IEEEpeerreviewmaketitle

\section{Introduction}
Modern computing systems, from the micro-scale of network-on-chip architectures \cite{marculescu2009outstanding} to large-scale server farms and cloud computing platforms \cite{celik2018wireless}, rely on the availability of effective communication links in order to operate over distributed architectures. 
Communication in distributed systems is required not only for the input/output procedures, but also for \emph{shuffling} data among the distributed system elements \cite{li2018fundamental}. 
As a result, the time and resources needed for communication can become the bottleneck in the operation of a distributed computing system, and the overall performance should account for the cost of both computing and communication. 

While distributed computing platforms typically assume the availability of reliable and dedicated connections among the processors, a number of emerging scenarios are characterized by wireless inter-processor links. 
These include wireless data centers \cite{celik2018wireless}, in which the processors are conventional servers, and federated learning \cite{konevcny2016federated, park2018wireless}, in which computing is carried out collaboratively by mobile devices. 
In the presence of wireless links, the design of the system should not only account for standard impairments related to computing, such as straggling processors \cite{lee2018speeding}, but also for issues arising from wireless transmissions, such as interference and imperfect Channel State Information (CSI). 

Under the assumption that communication is ideal, recent work has demonstrated the role of coding of the input data in mitigating the impact of stragglers \cite{lee2018speeding, dutta2018optimal}, as well as of coding of the output data in reducing the communication load for Map-Shuffle-Reduce systems \cite{li2018fundamental}.  
Assuming ideal computation (i.e., no stragglers) the impact of wireless interference in a Map-Shuffle-Reduce system with output data coding was studied in \cite{li2018wireless} and \cite{ha2018wireless}, with the former assuming perfect CSI and the latter accounting for imperfect CSI. 
These works adopt the communication delay as the performance criterion of interest.

In this work, we study for the first time the impact of both straggling processors and wireless communication impairments, such as interference and imperfect CSI, in the wireless distributed or federated system illustrated in Fig. \ref{fig_system}.
The total latency of the system, obtained as the sum of computing and communication delays, is investigated by taking an information-theoretic approach based on a high-Signal-to-Noise ratio (SNR) approximation of the communication delay \cite{sengupta2017}.
Under this metric, computation and communication protocols are proposed for the federated computation of multivariate polynomial functions based on Lagrange encoding of the data \cite{yu2018lagrange} and different shuffling communication strategies, namely coded multicasting \cite{li2016unified} and cooperative transmission \cite{li2018wireless}.
The analysis reveals the interplay between distributed computing, coding, and cooperative or coordinated transmission.

\textbf{Notation}: For any integer $P$ and $J$, we define the set $[P] \triangleq \{1,2,\cdots,P\}$, and the set $\{A_{j}\}_{j=1}^{J} \triangleq \{A_{1}, \cdots, A_{J}\}$.
We define $|\mathcal{A}|$ as the cardinality of set $\mathcal{A}$.
We also define the symbol $\doteq$ to denote an exponential equality: we write $f(P) \doteq P^{\alpha}$ if $\lim_{P \rightarrow \infty} \log(f(P))/\log(P) = \alpha$ holds.
Matrices and vectors are denoted by upper-case and lower-case bold fonts, respectively.

\section{System Model and Operation}
\label{sec:sm}

\subsection{System Model}
As illustrated in Fig. 1, we consider a distributed computing system in which $K$ full-duplex capable devices, or nodes, communicate over a shared wireless channel in order to cooperatively compute $N$ functions $\mathcal{F} = \{f_{n}\}_{n=1}^{N}$ over a data set $\mathcal{A} = \{\mathbf{a}_{i}\}_{i=1}^{m}$.
Function computation is a key step in many applications, including in distributed learning systems such as federated learning \cite{konevcny2016federated}.
We assume that distributed computation follows the Map-Shuffle-Reduce framework \cite{li2018fundamental, yan2018storage}.
Devices generally have limited storage capacity and different random online execution times for local computations.
A network controller holds the data set $\mathcal{A}$ and can communicate to the devices via out-of-band link.
Each data point $\mathbf{a}_{i}$ in $\mathcal{A}$ is from a vector space $\mathbb{V}$ over a sufficient large field $\mathbb{F}$.
Each function $f_{n}: \mathbb{V} \rightarrow \mathbb{U}$ takes values in a vector space $\mathbb{U}$ over field $\mathbb{F}_{2^{L}}$, and is assumed to be a multivariate polynomial of maximum degree $d$.
This class of functions includes standard tensor operations used in learning algorithms \cite{yu2018lagrange}.
For each function $f_{n}$, the output $\mathbf{y}_{n}$ over the data set is given as
\begin{equation}
\label{eq:yn}
\mathbf{y}_{n} = \{f_{n}(\mathbf{a}_{1}), \cdots, f_{n}(\mathbf{a}_{m})\}.
\end{equation}
The storage and processing capacity of each of the $K$ devices equals a fraction $\mu$ of the data set $\mathcal{A}$, with $\mu \in [1/K,1]$ being the fractional storage capacity.

The network controller can encode the data set $\mathcal{A}$ before communicating with the devices.
An $(m',m)$ linear code yields the coded data set $\mathcal{C} = \{\mathbf{c}_{i}\}_{i=1}^{m'}$,
defined as 
\begin{equation}
\label{eq:coded}
\mathbf{C} = [\mathbf{c}_{1}^{T}, \cdots, \mathbf{c}_{m'}^{T}]^{T} = \mathbf{G}\mathbf{A},
\end{equation}
where $\mathbf{A} = [\mathbf{a}_{1}, \cdots, \mathbf{a}_{m}]^{T}$ is the data matrix, and $\mathbf{G} \in \mathbb{F}^{m' \times m}$ is an encoding matrix, with integer $m' > m$. 
Each device $k$ can store and process up to $\mu m$ coded data rows, and we define as $\mathcal{C}_{k} \subseteq \mathcal{C}$ for $k \in [K]$ the subsets of rows of matrix $\mathbf{C}$ available at each device $k$, with $|\mathcal{C}_{k}| \leq \mu m$.
This implies that we can set $m' \leq \mu mK$ without loss of generality.

We assume that the channel between the devices is flat fading, so that the received signal at device $k$ is given as
\begin{equation}
\label{eq:channel}
y_{k} = \sum_{i \in [K]} h_{i,k}x_{i} + n_{k},    
\end{equation}
where $h_{i,k} \sim \mathcal{CN}(0,1)$ is the complex channel coefficient between device $i$ and device $k$, for $i$, $k \in [K]$; $x_{i}$ is the transmitted signal from device $i$ with power constraint $\mathbb{E}[|x_{i}|^{2}] \leq P$; and $n_{k} \sim \mathcal{CN}(0,1)$ is the additive Gaussian noise at device $k$.
As in \cite{li2018wireless, ha2018wireless}, each device is capable of full-duplex communication, i.e., each device can transmit and receive simultaneously.

Each device $k$ estimates the channels $\{h_{i,k}\}$ from all other devices in a training phase, and delivers the estimated CSI to the network controller.
The transmission schedule and beamforming vectors for all devices are designed by the network controller based on the received CSI and are transmitted to the devices.
Since there is a delay caused by processing and transmission, the CSI available at the network controller is assumed to be noisy and outdated with respect to the actual channel coefficients.
We model the remaining error between the outdated CSI $\{\hat{h}_{i,k}\}$ and the actual CSI $\{h_{i,k}\}$ as 
\begin{equation}
\mathbb{E}[|h_{i,k} - \hat{h}_{i,k}|^{2}] \doteq P^{-\alpha},
\end{equation}
for some $\alpha \geq 0$.
This model has been widely adopted for analyzing the imperfect CSI in the high-SNR regime (see, e.g., \cite{ImCSI}).
In this regime, the case $\alpha = 0$ yields that the network controller has no CSI, while the case $\alpha = 1$ implies that there is a negligible CSI error.
In contrast to the CSI available at the network controller, we assume that the CSI available at the receiver side during transmission between devices is accurate, which can be ensured by adding pilot symbols to each transmitted packet.

\subsection{Map-Shuffle-Reduce Protocol}
The three phases of operation of the system are as follows.
\begin{figure}[!t]
\centering
\includegraphics[width=4in]{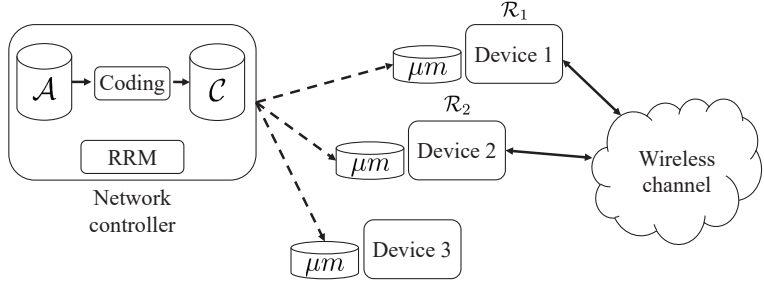}
\caption{Wireless Map-Reduce federated computing system with $K=3$ full-duplex devices, each able to store a (possibly coded) fraction $\mu m$ of the input data. In this example, Device 3 is a straggler in the Map phase, and the computing outputs are assigned to Device 1 and Device 2. Radio Resource Management (RRM) refers to scheduling and beamforming design carried out at the network controller.}
\label{fig_system}
\end{figure}

\emph{Map phase}: In the Map phase, the network controller sends a subset $\mathcal{C}_{k}$ of coded rows from \eqref{eq:coded} to each device.
Each device $k$ then computes the function in $\mathcal{F}$ for all stored data $\mathbf{c} \in \mathcal{C}_{k}$.
This produces a set of Intermediate Values (IVs) computed by device $k$ as $\mathcal{I}_{k} = \{f_{n}(\mathbf{c}) : f_{n} \in \mathcal{F}, \mathbf{c} \in \mathcal{C}_{k}\}$.
The computation time of the devices is random, and hence some devices may be straggling as compared to the others.
After the computation of the respective IVs is completed at $q \leq K$ devices, where $q$ is a predetermined parameter to be designed, the Map phase is considered to be over.

\emph{Shuffle phase}: Define as $\mathcal{Q} \subseteq [K]$ as the set of $q$ non-straggling devices, with $|\mathcal{Q}| = q$.
Each non-straggling device $k$ is assigned $N/q$ arbitrary disjoint output functions $\{f_{n} : n \in \mathcal{R}_{k}\}$, for some subset $\mathcal{R}_{k} \in [1,N]$ of the $N$ output $\{\mathbf{y}_{n}\}_{n=1}^{N}$, with $\bigcup_{k \in \mathcal{Q}} \mathcal{R}_{k} = [N]$.
In the Shuffle phase, the devices in set $\mathcal{Q}$ communicate on the wireless channel \eqref{eq:channel} in order to enable each device $q \in \mathcal{Q}$ to compute the output $\mathbf{y}_{n}$ for $n \in \mathcal{R}_{k}$ by exchanging locally computed IVs.

\emph{Reduce phase}: In the Reduce phase, each device $k$ computes the assigned outputs $\{\mathbf{y}_{n} : n \in \mathcal{R}_{k}\}$ based on the locally computed IVs in set $\mathcal{I}_{k}$ and the IVs received from other devices in the Shuffle phase.
Each device $k$ transmits the computed outputs $\{\mathbf{y}_{n} : n \in \mathcal{R}_{k}\}$ to the network controller.

\subsection{Performance Criterion}
As the performance criterion of interest, we define the total average delay $\delta_{T}$ as the sum of the average computation time $\delta_{M}$ during the Map phase, the communication time $\delta_{S}$ during the Shuffle phase, and the computation time $\delta_{R}$ during the Reduce phase.
We now discuss each term in turn.

\emph{Map phase}: The Map phase delay is defined as the average time required for first $q$ devices to complete their IV computations in the respective subset $\mathcal{I}_{k}$.
As in \cite{li2016unified}, we assume that the time needed for computing the subset $\mathcal{I}_{k}$ of IVs at each device $k$ has a shifted exponential distribution with shift and average proportional to the number $\mu m$ of processed coded data points.
We normalize the average Map phase delay by the average time needed to compute IVs over $Nm$ data points, i.e., over the entire input data set $\mathcal{A}$, at a device.
The resulting normalized Map phase delay $\delta_{M}$ per input data is given as \cite{li2016unified}   
\begin{equation}
\label{eq:mpd}
\delta_{M}(\mu, q) = \frac{\mu}{2} \left(1+\sum_{j=K - q + 1}^{K}\frac{1}{j}\right).
\end{equation}
The Map phase delay $\delta_{M}(\mu,q)$ is an increasing function of $\mu$ and $q$.

\emph{Shuffle phase}: We measure the Shuffle phase communication delay in the high-signal-to-noise ratio (SNR) regime by following \cite{sengupta2017}, \cite{ li2018wireless, ha2018wireless}.
As detailed in \cite{sengupta2017}, this allows us to focus on the impact of mutual interference the wireless channel while obtaining a tractable latency metric.
To elaborate, define as $T$ the total time required for each device $k \in \mathcal{Q}$ to receive all the IVs needed to reduce the assigned functions.
In order to measure this quantity in the high-SNR domain, we normalize $T$ by the time $NmL/\log(P)$ needed in the high-SNR regime to communicate the $Nm$ outputs $f_{n}(\mathbf{a}_{i})$, for $n=1,\cdots,N$ and $i=1,\cdots,m$, computed on the entire data set to a device in the absence of mutual interference (i.e., with high-SNR rate $\log$(SNR)).
The resulting average Shuffle phase delay per input data is given as  
\begin{equation}
\label{eq:sd}
\delta_{S}(\mu, q) = \lim_{P \rightarrow \infty}\frac{\mathbb{E}[T]}{NmL/\log(P)}.
\end{equation}

\emph{Reduce phase}: The Reduce phase delay is defined as the time required for the $q$ devices in the set $\mathcal{Q}$ to compute the assigned outputs and transmit their outputs to a central unit.
Each device $k$ in set $\mathcal{Q}$ should compute $|\mathcal{R}_{k}|=N/q$ outputs, and hence the corresponding average computing delay at each device generally decreases with $q$.

In this work, we assume that the Reduce phase delay is negligible as compared to the Map and Shuffle phase delays, which is generally the case when the input space $\mathbb{V}$ is much larger than the output space $\mathbb{F}_{2^{L}}$.
The analysis can be easily extended to account also for the Reduce phase delay.
Accordingly, we define the sum of the Map phase delay and Shuffle phase delay as the total delay as 
\begin{equation}
\label{eq:td}
\delta_{T} = \gamma\delta_{M} + \delta_{S}.
\end{equation}
In \eqref{eq:td}, parameter $\gamma$ equals the ratio between the average time (in seconds) needed to compute one bit of the input data at a device and the average time (in seconds) needed to transmit one bit in an interference-free channel.
In practice, this parameter can range, e.g., from 0.1 for a powerful device such as laptop to 10 for a smart phone (see, e.g., \cite{mao2016dynamic, zhang2018energy}). 

\section{Preliminaries}
\label{sec:BS}
\begin{figure}
\centering
\begin{subfigure}[b]{5in}
\centering
  \includegraphics[width=4in]{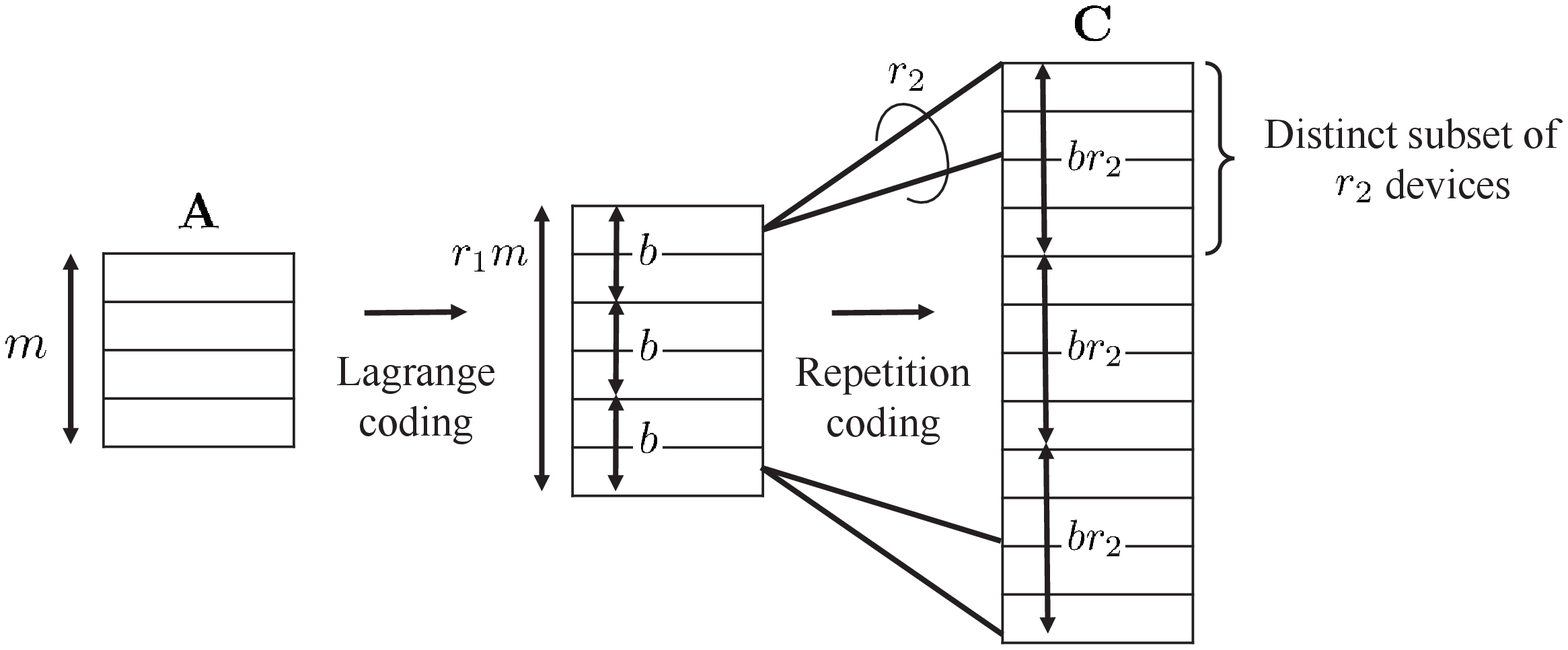}
  \caption{}
  \label{fig:Ng1} 
\end{subfigure}
\begin{subfigure}[b]{5in}
\centering
  \includegraphics[width=2in]{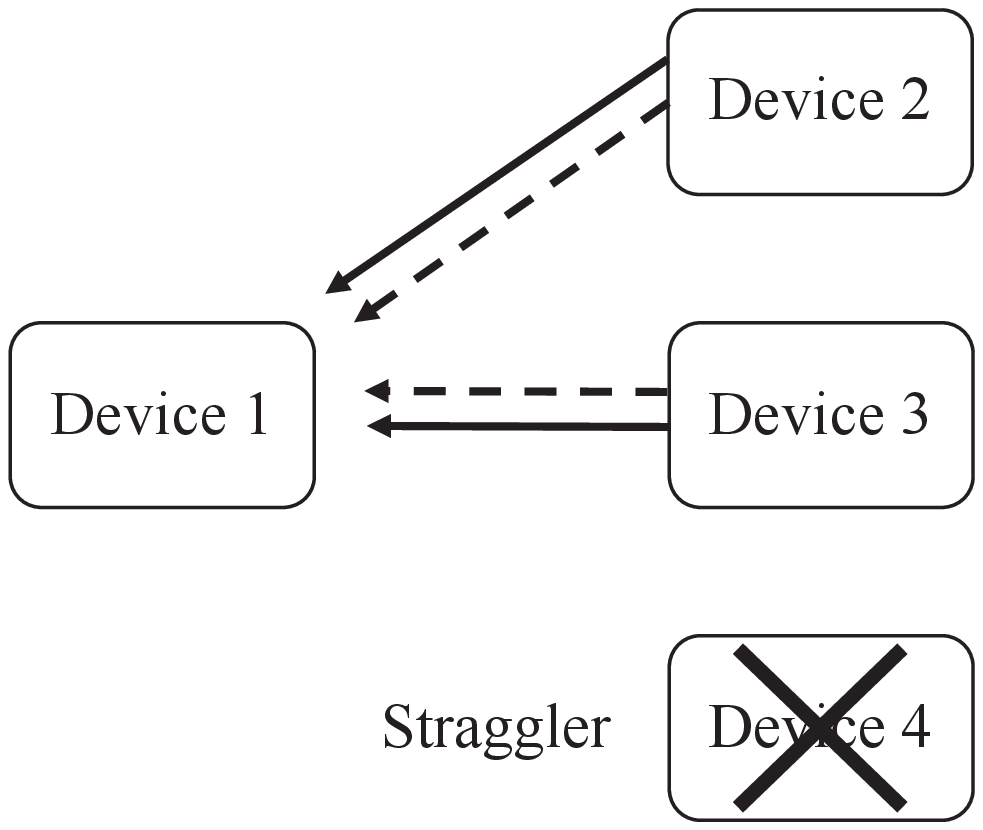}
  \caption{}
  \label{fig:Ng2}
\end{subfigure}
\caption[]{(a) Map phase based on concatenated coding \cite{zhang2018improved}; and (b) Shuffle phase from the viewpoint of Device 1: Device 1 needs to receive a number of IVs available only at Device 2 and 3 (solid arrows) respectively, and some available at both Device 2 and 3 (dashed arrows).}
\label{fig_system2}
\end{figure}

In order to provide the necessary background, in this section, we first review a useful result concerning Lagrange coding for master-slave systems from \cite{yu2018lagrange}, and then we present a generalization of a result from \cite{zhang2018improved} on concatenated coding for Map-Shuffle-Reduce systems.

In \cite{yu2018lagrange}, Lagrange coding is introduced as a way to ensure robustness to stragglers in a master-slave computing system aimed at evaluating multivariate polynomial functions.
The Lagrange linear $(m'=r_{1}m,m)$ code has the following key property.

\emph{Lemma 1}: For a multivariate polynomial function of degree $d$, given an $(r_{1}m,m)$ Lagrange linear code \eqref{eq:coded} producing a set $\mathcal{C}$ of $r_{1}m \geq md - 1$ coded data points, it is possible to recover the $m$ outputs in $\mathbf{y} = [f(\mathbf{a}_{1} \cdots f(\mathbf{a}_{m})]$ from any subset $m^{*}$ of IVs $f(\mathbf{c})$ with $\mathbf{c} \in \mathcal{C}$, where
\begin{equation}
\label{eq:crt}
m^{*} = (m-1)d' + 1,
\end{equation}
and we have $d' = 1$ if $r_{1} = 1$ (and hence no coding is used) and $d' = d$ if $r_{1} > 1$.
The quantity $m^{*}$ is known as the recovery threshold of Lagrange coding. 

We now consider a generalization of the concatenated coding scheme proposed in \cite{zhang2018improved} for linear functions.
The extension applies to the more general class of multivariate polynomial functions.
As seen in Fig. \ref{fig_system2}(a), the linear $(r_{1}r_{2}m,m)$ coding scheme concatenates a Lagrange code with redundancy $r_{1} \in \mathbb{R}$ and a repetition code with redundancy $r_{2} \in \mathbb{N}^{+}$.
For a distributed computing system with devices having the same storage and processing constraints as for the system in Fig. \ref{fig_system}, we have the condition $m' \leq \mu mK$, which implies the inequality $r_{1}r_{2} \leq \mu K$.
Furthermore, each Lagrange encoded row is stored and processed at $r_{2}$ devices.
To define the assignment of the $r_{1}m$ Lagrange coded rows to the devices, i.e., to define the subsets $\mathcal{C}_{k} \subseteq \mathcal{C}$, we write the number of Lagrange coded rows as $r_{1}m = \binom{K}{r_{2}}b$ for some integer $b \in \mathbb{N}^{+}$ under the assumption that $m$ is large enough.
The $r_{1}m$ Lagrange coded rows are then divided into $\binom{K}{r_{2}}$ batches of size $b$, with each batch stored at a disjoint subset $\mathcal{K} \subseteq [K]$ of $r_{2}$ devices (see Fig. \ref{fig_system2}(a) for an illustration).

From \emph{Lemma 1}, in order for any set of $q$ non-straggling devices to collectively have enough information to recover all outputs \eqref{eq:yn} for a given function $f(\cdot)$, the $q$ devices should compute IVs evaluated on at least $m^{*}$ distinct Lagrange encoded data points.
Extending \cite[\emph{Proposition 1}]{zhang2018improved}, this condition is satisfied if the following inequalities hold:
\begin{subequations}
\vspace{-3mm}
\label{eqcon}
\begin{align}
\label{eqcon1}
&r_{2} > K-q, \hspace{1mm} \textrm{if} \hspace{1mm}r_{1} = 1, \hspace{1mm} \textrm{and}\\
\label{eqcon2}
&\binom{K}{r_{2}} - \binom{K-q}{r_{2}} \geq \frac{m^{*}}{r_{1}m} {\binom{K}{r_{2}}},  \hspace{1mm} \textrm{if} \hspace{1mm}r_{1} > 1.
\end{align}
\end{subequations}
This can be briefly proved as follows.
If $r_{1}=1$, Lagrange coding is not used, and the repetition factor $r_{2}$ should be larger than the number of straggling devices, so that every row of data set is stored at least one non-straggling device, i.e., $r_{2} > K - q$.
Instead, if $r_{1} > 1$, the number of Lagrange coded rows stored exclusively at any subset of $K-q$ straggling devices is $b\binom{K-q}{r_{2}}$.
The number of IVs evaluated on distinct Lagrange coded rows at the non-straggling devices is hence $b\big(\binom{K}{r_{2}} - \binom{K-q}{r_{2}}\big)$.
Imposing that this number be larger than $m^{*}$ yields the inequality in \eqref{eqcon2}.
We finally note that condition \eqref{eqcon} implies the lower bound $q \geq q_{\min} = \min(\lceil ((m-1)d+1)/\mu m\rceil, K- \lfloor \mu K \rfloor + 1)$ on the number $q$ of non-straggling devices.

While condition \eqref{eqcon} ensures that all non-straggling devices have collectively enough information to recover a desired output, the non-straggling devices need to exchange IVs in the Shuffle phase on the wireless channel so as to enable a successful Reduce phase for all functions in $\mathcal{F}$.
This is discussed in the next section.

\section{Shuffling Schemes}
\label{sec:SC}
In this section, we propose three shuffling schemes that enable the successful completion of the Reduce phase for the concatenated coding strategy described in Sec. \ref{sec:BS}.
For all schemes, by \emph{Lemma 1}, at the end of the Shuffle phase, each device $k$ needs to have $m^{*}$ distinct IVs for the reconstruction of each of the $N/q$ assigned outputs $\{\mathbf{y}_{n} : n \in \mathcal{R}_{k}\}$ in the Reduce phase. 
Since each device $k$ computes $|\mathcal{C}_{k}|$ distinct IVs for each assigned output $\mathbf{y}_{n}$, $m^{*} - |\mathcal{C}_{k}|$ IVs per function, for a total of $(m^{*} - |\mathcal{C}_{k}|)N/q$ IVs, need to be received from the other non-straggling devices in the set $\mathcal{Q}$.
Thanks to the repetition code, each IV desired by a device $k$ is generally available to a number of other devices, which we refer to as the computational redundancy or \textit{multiplicity} of the IV.

For each IV, the multiplicity is no larger than $r_{2}$ and can be seen to range in the interval $[s_{\min}:s_{\max}]$, where $s_{\min} = \max(r_{2}-(K-q),0)$ and $s_{\max} = \min(q-1, r_{2})$ \cite{zhang2018improved} (see Fig. \ref{fig_system2}(b) for an illustration). 
Furthermore, for each multiplicity $j\in[s_{\min}:s_{\max}]$, the number of IVs per output is given as
\begin{equation}
B_j=b\binom{q-1}{j}\binom{K-q}{r_{2}-j}.
\end{equation}
This is because for every one of the $\binom{q-1}{j}$ subsets of size $j$ of other non-straggling devices, there are $\binom{K-q}{r_{2}-j}$ subsets of size $r_{2}-j$ of straggling devices that share the same IVs.

All schemes deliver IVs in order of decreasing multiplicity since, as we will see, a larger multiplicity implies a lower contribution to the Shuffle delay.
Given that the total number of IVs per output to be shuffled is $m^{*}-|\mathcal{C}_{k}|$, this implies that all IVs with multiplicity ranging from $s_{\max}$ down to $s_{q}$ with $s_{q} = \inf \{s : \sum_{j=s}^{s_{\max}}B_{j} \leq m^*- |\mathcal{C}_k|\}$ are exchanged in full, while the rest of the $m^*-|\mathcal{C}_k|-\sum_{j=s_{q}}^{s_{\max}}B_{j}$ IVs to be exchanged have multiplicity $s_{q}-1$.


\subsection{Coded Multicasting}
\label{subsec:CS}
In \cite{li2016unified}, a coded multicasting transmission scheme is introduced for the Shuffle phase for the case of linear function computation, ideal multicasting communication, and no stragglers.
In \cite{zhang2018improved}, the analysis is extended to the scenario with stragglers.
Here we generalize the scheme and the analysis to the scenario at hand.
Coded multicasting leverages computational redundancy, that is, a multiplicity larger than one, by scheduling a sequence of one-to-many multicasting transmissions that are simultaneously useful to more devices.


As discussed, IVs are shuffled sequentially in order of decreasing multiplicity $j$ from $s_{\max}$ to $s_{q}-1$.
Furthermore, in group $j$, $B_{j}$ IVs per output have to be exchanged, while $m^{*}-|\mathcal{C}_{k}|-\sum_{j=s_{q}}^{s_{\max}}B_{j}$ IVs per function are exchanged for group $s_{q}-1$. 
Devices transmit in turn by serving $j$ other users simultaneously via coded multicasting, whereby the $j$ IVs are XORed and decoding leverages the available IVs as side information \cite{li2016unified}.

\emph{Proposition 1}: For storage capacity $\mu \in [1/K,1]$, degree $d$ of the multivariate polynomial functions, number of distributed devices $K$, and number $q \in [q_{\min} : K]$ of non-straggling devices, the Shuffle phase delay \eqref{eq:sd} of coded multicasting is given as 
\begin{equation}
\label{eq:CM}
\delta_{S}^{CM}(\mu,q) = \min_{\substack{r_{1} \in [1:\mu K]\\r_{2} \in [1:\lfloor\mu K\rfloor]}}
 \sum_{j=s_{q}}^{s_{\max}}\frac{B_{j}}{mj}+ \frac{m^{*} - |\mathcal{C}_{k}|-\sum_{j=s_{q}}^{s_{\max}}B_{j}}{m(s_{q}-1)},
\end{equation}
where the minimization is subject to constraints \eqref{eqcon}.

\emph{Proof}: The proof follows immediately by noting that the first sum is the normalized delay \eqref{eq:sd} for the transmission of $B_{j}$ IVs given the coded multicasting gain of $j$, while the second term corresponds to the transmission of the remaining IVs.
Minimization is carried out over the parameters $(r_{1},r_{2})$ of the concatenated code.
See also \cite{zhang2018improved} for further details.

\emph{Remark 1}: When the degree of function is $d=1$, the Shuffle phase delay \eqref{eq:CM} coincides with the communication load derived in \cite[\emph{Proposition 2}]{zhang2018improved} normalized by $N$.
xhxmsja
\subsection{One-Shot Linear Precoding}
\label{subsec:OS}
In \cite{li2018wireless}, a one-shot linear Zero-Forcing (ZF) precoding scheme is proposed that applies to linear functions for the case of no stragglers and of a wireless channel with perfect CSI.
Here we extend the analysis to more general multivariate polynomial functions, imperfect CSI, and straggling devices.

Unlike coded multicasting, this scheme leverages computational redundancy by enabling cooperative simultaneous transmission by clusters of devices that have computed the same IVs.
Specifically, for each group of IVs with multiplicity $j$, $2j$ devices transmit simultaneously at any given time.
The $2j$ devices are split into two clusters of size $j$, with the property that a device in one cluster has computed a required IV for all the devices of the other cluster.
Thanks to full-duplex communication, the two clusters can transmit simultaneously to one another.
Furthermore, the devices in each cluster apply ZF precoding in order to communicate to the devices in the other cluster without mutual interference.
Interference from other devices in the same cluster caused by full-duplex transmission can be removed, since each device knows the IVs sent by other devices in the same cluster, 
This discussion applies to the case $q \geq 2j$ and for the case of $q < 2j$, $q$ IVs can be simultaneously transmitted in a similar way \cite{li2018wireless}.

As a result, with one-shot linear precoding, $\min(q,2j)$ IVs are delivered in a single transmission without mutual interference.
In the presence of imperfect CSI, the high-SNR transmission rate of ZF precoding is given as $\alpha\log$(SNR), and is hence decreasing with the CSI accuracy parameter $\alpha$ \cite{ImCSI}.

\emph{Proposition 2}: For storage capacity $\mu \in [1/K,1]$, degree $d$ of the multivariate polynomial functions, number of distributed devices $K$, and number $q \in [q_{\min} : K]$ of non-straggling devices, the Shuffle phase delay \eqref{eq:sd} of one-shot cooperative linear precoding is given as
\begin{equation}
\label{eq:ZF}
\delta_{S}^{ZF}(\mu,q) = \min_{\substack{r_{1} \in [1:\mu K]\\r_{2} \in [1:\lfloor\mu K\rfloor]}} \sum_{j=s_{q}}^{s_{\max}}\frac{B_{j}}{m\alpha\min(q,2j)} + \frac{m^{*} - |\mathcal{C}_{k}|-\sum_{j=s_{q}}^{s_{\max}}B_{j}}{m\alpha\min(q,2(s_{q}-1))},
\end{equation}
where the minimization is subject to constraints \eqref{eqcon}.

\emph{Proof}: The proposition follows directly in the same way as for \emph{Proposition 1} based on the discussion above.

\emph{Remark 2}: When $d=1$ and $r_{1}=1$, the Shuffle phase delay \eqref{eq:ZF} coincide with the communication load derived in \cite[\emph{Theorem 1}]{li2018wireless}.

\subsection{Superposition Coding}
\label{subsec:SC}
In \cite{ha2018wireless}, a superposition-coding based transmission is introduced in order to reduce the Shuffle phase delay of the ZF-based scheme in \cite{li2018wireless} in the presence of imperfect CSI and with no stragglers.
Here we extend the approach to multivariate polynomial functions and to account for possible stragglers.

As in the ZF scheme described above, for each group $j$, $2j$ devices transmit simultaneously.
The difference in that, with superposition coding, the ZF-precoded signals are sent with the smaller power $P^{\alpha}$, and one of $2j$ active devices superimposes on the ZF-precoded signal a coded multicasting signal.
This signal is intended for $j$ other non-straggling devices and is sent with the power $P-P^{\alpha}$.
Furthermore, each device first decodes the coded multicasting signal by treating ZF-precoded signals as noise, and then decodes the ZF-precoded signals by using Successive Interference Cancellation (SIC).
This scheme deliver $j$ IVs by coded multicasting with high-SNR rate $\log((P-P^{\alpha})/P^{\alpha}) \doteq (1-\alpha)\log(P)$, and $2j$ IVs by one-shot linear precoding with rate $\alpha\log(P)$ in a single transmission.
The transmission rate for coded IV decreases with $1-\alpha$ due to the interference from ZF-precoded signal. 

\emph{Proposition 3}: For storage capacity $\mu \in [1/K,1]$, degree $d$ of the multivariate polynomial functions, number of distributed devices $K$, and number $q \in [q_{\min} : K]$ of non-straggling devices, the Shuffle phase delay \eqref{eq:sd} of superposition coding is given as
\begin{equation}
\label{eq:SC}
\delta_{S}^{SC}(\mu,q) =  \min_{\substack{r_{1} \in [1:\mu K]\\r_{2} \in [1:\lfloor\mu K\rfloor]}} \sum_{j=s_{q}}^{s_{\max}}\frac{B_{j}}{m[(1-\alpha)j + \alpha \min(q,2j)]} + \frac{m^{*} - |\mathcal{C}_{k}|-\sum_{j=s_{q}}^{s_{\max}}B_{j}}{m[(1-\alpha)(s_{q}-1) + \alpha \min(q,2(s_{q}-1))]},   
\end{equation}
where the minimization is subject to constraints \eqref{eqcon}.

\emph{Proof}: For each group $j$ of IVs, the effective high-SNR transmission rate is given as $[(1-\alpha)j + \alpha\min(q,2j)]\log$(SNR) due to the simultaneous transmission of ZF-precoded and coded multicasting signals.
The proof is completed as for \emph{Proposition 1} and \emph{Proposition 2}.

\emph{Remark 3}: If $r_{1}=1$ and there are no stragglers, i.e., $q=K$, the Shuffle phase delay \eqref{eq:SC} is the same as the communication load derived in \cite[\emph{Proposition 3}]{ha2018wireless}.
Furthermore, it can be proved as in \cite{ha2018wireless} that the Shuffle phase delay \eqref{eq:SC} is no larger than both \eqref{eq:CM} and \eqref{eq:ZF}.

\section{Numerical results}
In this section, we compare the performance of the three schemes described in Sec. \ref{sec:BS} via a numerical experiment by considering a wireless distributed computing system with $K = 30$ distributed devices with a fractional storage capacity $\mu = 1/2$ that aims at computing $N = 120$ polynomial functions with degree $d \geq 1$ over a data set $\mathcal{A}$ with $m = 600$ rows.

\begin{figure}[!t]
\centering
\includegraphics[width=4in]{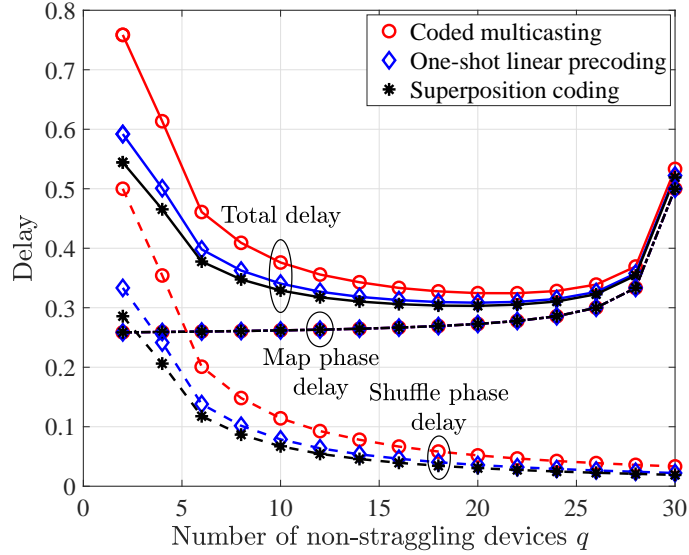}
\caption{Map phase delay $\delta_{M}$ in \eqref{eq:mpd}, Shuffle phase delay $\delta_{S}$ derived in Sec. \ref{sec:SC}, and total delay $\delta_{T}$ in \eqref{eq:td} as a function of the number of non-straggling devices $q$ for the different schemes, with $d = 1$, $\gamma = 1$ and $\alpha = 0.75$.}
\label{fig_result1}
\end{figure}

We first plot separately the Map phase delay $\delta_{M}$ in \eqref{eq:mpd}, the Shuffle phase delay $\delta_{S}$ derived in Sec. \ref{sec:SC} and the total delay $\delta_{T}$ in \eqref{eq:td} as a function of the number of non-straggling devices $q$, where we have set degree $d = 1$, computation-to-communication delay ratio $\gamma = 1$, and CSI accuracy $\alpha = 0.75$.
The Map phase delay of all schemes is the same and is seen to increases with $q$, since a larger $q$ implies waiting for more devices to complete their computations.
In contrast, the Shuffle phase delay of all schemes decreases with $q$, since a larger $q$ allows to increase either the multicasting or the cooperation opportunities.
Accordingly, there is an optimal value of $q$ that minimizes the total delay $\delta_{T}$.
We also observe that the superposition coding scheme outperforms both multicasting and cooperative transmission schemes for all number of non-straggling devices $q$.

\begin{figure}[!t]
\centering
\includegraphics[width=4in]{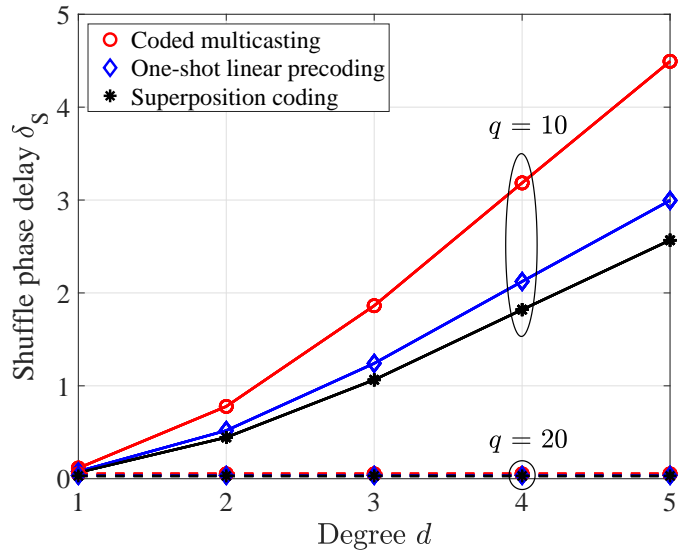}
\caption{Shuffle phase delay $\delta_{S}$ derived in Sec. \ref{sec:SC} as a function of the degree $d$ of the multivariate polynomial functions to be computed for the different Shuffling schemes, with $\gamma = 1$, $\alpha = 0.75$, and $q=10,20$.}
\label{fig_result2}
\end{figure}

The impact of the degree $d$ on the Shuffle phase delay $\delta_{S}$ is shown in Fig. \ref{fig_result2}, with  $\gamma = 1$, $\alpha = 0.75$, and different values of $q$.
First, we observe that the Shuffle phase delay of all schemes increase with $d$ when $q=10$.
This is because the recovery threshold $m^{*}$ in \eqref{eq:crt} of Lagrange coding increases proportionally to the degree $d$.
In contrast, the Shuffle phase delay of all schemes is independent of $d$ when $q=20$.
In fact, in this case, there are enough non-straggling devices to receive all required IVs by using only the redundancy of the repetition code, and the minimum Shuffle phase delay is obtained with $r_{1}=1$.

\begin{figure}[!t]
\centering
\includegraphics[width=4.5in]{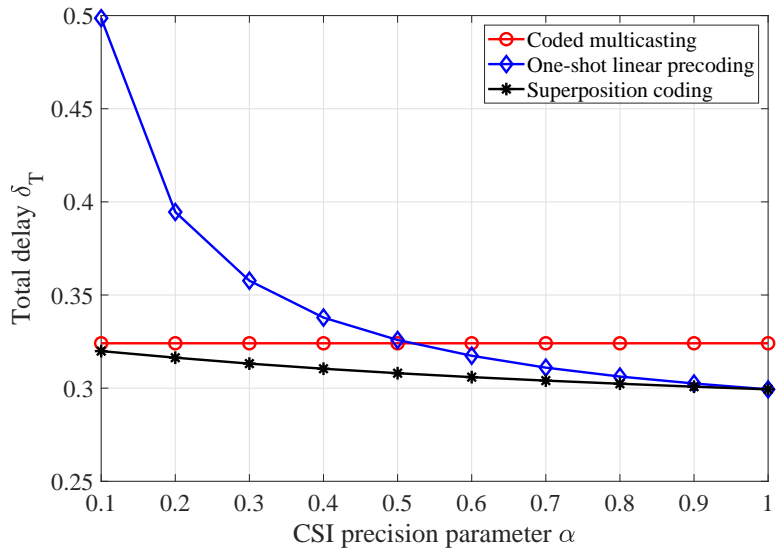}
\caption{Total delay $\delta_{T}$ in \eqref{eq:td} as a function of the CSI precision parameter $\alpha$ for the different Shuffling schemes, with $d = 1$, $\gamma = 1$.}
\label{fig_result3}
\end{figure}

Finally, in Fig. \ref{fig_result3}, we investigate the total delay $\delta_{T}$ dependence of function of the CSI precision parameter $\alpha$, for $d = 1$ and $\gamma = 1$.
The curves in Fig. \ref{fig_result3} are plotted for the optimal numbers of non-straggling devices $q$ that minimizes the total delay $\delta_{T}$ for each $\alpha$.
The figure confirms that superposition coding outperforms both coded multicasting and cooperative transmission schemes and that, as the CSI precision parameter $\alpha$ increases, one-shot linear precoding scheme tends to yield a lower latency due to the improved accuracy of ZF precoding.
When $\alpha = 1$, the total delay for superposition coding scheme coincides with the total delay for one-shot linear precoding.
In contrast, when $\alpha = 0$, the total delay for superposition coding scheme coincides with the total delay for coded multicasting scheme.

\section{Conclusions}
In this paper, we have studied for the first time a wireless federated computing system based on the Map-Shuffle-Reduce framework in the presence of straggling device and imperfect CSI.
We have proposed a concatenated coding scheme that applies Lagrange coding and repetition coding along with coded multicasting or cooperative Shuffling communication strategies.
The high-SNR analysis of the total delay reveals the synergy between input data coding against stragglers and multicasting or cooperative transmission opportunities in the Shuffle phase.
We also demonstrated the advantages of a superposition-coding based scheme in the presence of imperfect CSI \cite{joudh2016ratesplitting}.

\section*{Acknowledgments}
The work of S. Ha and J. Kang was supported by the National Research Foundation of Korea (NRF) grant funded by the Korea government (MSIT) (No. 2017R1A2B2012698). J. Zhang and O. Simeone have received funding from the European Research Council (ERC) under the European Union’s Horizon 2020 Research and Innovation Programme (Grant Agreement No. 725731).

\ifCLASSOPTIONcaptionsoff
  \newpage
\fi
\bibliographystyle{IEEEtran}
\bibliography{refs}






\end{document}